\pdfoutput=1
\documentclass[journal]{IEEEtran}
\usepackage{subcaption}
\usepackage{graphicx}
\usepackage{mathtools}
\usepackage[latin1]{inputenc}
\usepackage{times,amsmath}
\usepackage{amsfonts}
\usepackage{pstool}
\usepackage{multirow}
\usepackage{multicol}
\usepackage{enumerate}
\usepackage{graphicx}
\usepackage{mathtools, cuted}
\usepackage{MnSymbol}
\usepackage{hyperref}
\usepackage{stfloats}
\usepackage{diagbox}
\usepackage[table]{xcolor}
\usepackage{algorithm,algorithmic}
\usepackage{bigints}
\usepackage{cite}
\usepackage{}

\begin{document}
\title{Your smartphone could act as a pulse-oximeter and as a single-lead ECG}
\author{

\IEEEauthorblockN{Ahsan Mehmood$^\ast$, Asma \ Sarauji$^\ast$, M.\ Mahboob\ Ur\ Rahman$^\ast$,  Tareq\ Y.\ Al-Naffouri$^\ast$  \\
$^ { \ast}$Department of Electrical Engineering, KAUST, Thuwal, KSA \\
$^\ast$\{ahsan.mehmood, asma.sarouji, muhammad.rahman, tareq.alnaffouri\}@kaust.edu.sa
}
}
\maketitle
\thispagestyle{plain}
\pagestyle{plain}

\begin{abstract}

In the post-covid19 era, every new wave of the pandemic causes an increased concern/interest among the masses to learn more about their state of well-being. Therefore, it is the need of the hour to come up with ubiquitous, low-cost, non-invasive tools for rapid and continuous monitoring of body vitals that reflect the status of one's overall health. In this backdrop, this work proposes a deep learning approach to turn a smartphone---the popular hand-held personal gadget---into a diagnostic tool to measure/monitor the three most important body vitals, i.e., pulse rate (PR), blood oxygen saturation level (aka SpO2), and respiratory rate (RR). Furthermore, we propose another method that could extract a single-lead electrocardiograph (ECG) of the subject. The proposed methods include the following core steps: subject records a small video of his/her fingertip by placing his/her finger on the rear camera of the smartphone, and the recorded video is pre-processed to extract the filtered and/or detrended video-photoplethysmography (vPPG) signal, which is then fed to custom-built convolutional neural networks (CNN), which eventually spit-out the vitals (PR, SpO2, and RR) as well as a single-lead ECG of the subject. To be precise, the contribution of this paper is two-fold: 1) estimation of the three body vitals (PR, SpO2, RR) from the vPPG data {using custom-built CNNs, vision transformer, and most importantly by CLIP model}; 2) a novel discrete cosine transform+feedforward neural network-based method that translates the recorded video-PPG signal to a single-lead ECG signal. The proposed method is anticipated to find its application in several use-case scenarios, e.g., remote healthcare, mobile health, fitness, and sports, etc. 
\end{abstract}
\begin{IEEEkeywords}
Pulse rate, SpO2, respiratory rate, photoplethysmography (PPG), pulse-oximeter, electrocardiograph, deep supervised learning, convolutional neural network, feedforward neural network, discrete cosine transform, vision transformer, CLIP. 
\end{IEEEkeywords}
\section{Introduction}

One flip-side of the Covid19 pandemic is that it has sparked great interest among people to stay informed about their overall well-being. Such interest in self-examination of one's own body vitals hikes whenever a new wave of covid19 strikes the world. By now, it is well known that the important body vitals deviate from their nominal values when one is infected with covid19  \cite{caruso2022effect, parizad2021effect } \footnote{{Throughout this paper, the term "vitals" is used to collectively refer to the pulse rate, blood oxygen saturation, respiratory rate, and occasionally blood pressure.}}. For example, the body temperature of a covid19 patient is often elevated ($100^{\circ}F$ or more). Additionally, a blood oxygen saturation level (aka SpO2) of less than 90\% and a respiratory rate of less than 15 and more than 25 typically also indicate a possible covid19 infection. Last, but not the least, an elevated pulse rate may also be witnessed as an occasional side-effect of covid19. Need not to say, but the body vitals have huge clinical significance other than covid19 detection as well. For example, the deviation of RR from nominal value could indicate the following: cardiac arrest \cite{chelluri2019respiratory}, respiratory dysfunction \cite{torsney2017respiratory}, pneumonia \cite{ginsburg2018systematic}, lungs cancer \cite{charlton2017breathing}. 

Thus, enabling masses to self-measure their body vitals in a rapid and non-invasive manner using low-cost, portable equipment (e.g., smartphones, smart watches, wristbands, etc.) is an objective of key importance, as suggested by the {world health organization} (WHO) \cite{chalabianloosmart}. To this end, note that there are currently more than 6 Billion active smartphone users, and this number will rise to 7 Billion by 2025 \cite{o'dea_23_2022}. Therefore, it is only logical to devise mechanisms that exploit the onboard sensors and computational capability of modern-day smartphones to help realize the patient-centric healthcare systems of tomorrow. 

In fact, there has been a recent surge of interest in smartphone-based estimation of body vitals, by extraction of various kinds of physiological signals. To this end, approaches that could extract photoplethysmography (PPG), electrocardiography (ECG), and phonocardiography (PCG) signals from the smartphone have been explored the most. PPG methods utilize the camera of the smartphone to measure the quantity of visible light reflected off the fingertip. A review of the PPG techniques and their potential applications could be found in \cite{tamura2019current}, \cite{castaneda2018review}. The ECG methods compute the electric potential difference across a pair of external electrodes mounted on the casing of the smartphone\cite{sagirova2021cuffless} (when one places his/her thumbs on the two electrodes). Finally, the PCG methods try to utilize the microphone of the smartphone to listen to faint sounds generated by the heart during each cardiac cycle. 

\begin{figure*}

    \centering
    \includegraphics[width=7in]{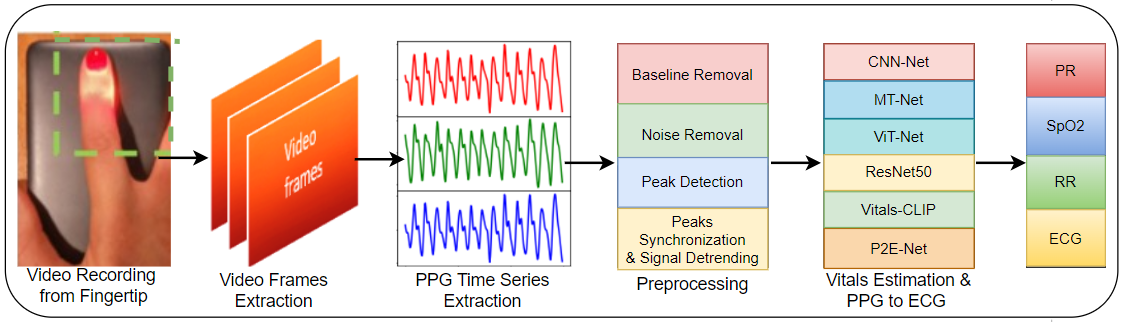}
    \caption{{A quick graphical summary of this work.} }
    \label{fig:1b }
    
\end{figure*}

In this pretext, we propose the utilization of deep learning techniques on PPG data derived from a smartphone to create a physiological monitor that can aid in the self-assessment of body vitals. Our study aims to evaluate the accuracy, robustness, and generalization capability of several deep learning models, including convolutional neural networks (CNNs), vision transformer (ViT), and CLIP model (that generates captions of given images), for the estimation of vitals and the reconstruction of electrocardiograms (ECGs). The core steps of our work are depicted in Fig. 1 and are elaborated on as follows:
\begin{itemize}
    \item \textit{Video Recording}: one needs to place his/her index finger on the rear camera of the smartphone in order to record a small video snippet of a small duration (say, 30 seconds).

    \item \textit{Video Preprocessing}: On RGB {(red, green, blue)} channels of video, pixel-averaging on each frame is done to get a PPG time series, and Wavelet transform is applied for denoising and removal of motion-induced artifacts. 
    \item \textit{Training and Testing of custom-built Neural Networks}: once the PPG time series is available, it is passed to a number of deep neural networks which eventually spit out the vitals (PR, SpO2, and RR) as well as a single-lead ECG of the subject. 
    
\end{itemize}

{\bf Outline.}
The rest of this paper is organized as follows. Section II describes selected related work. Section III discusses PPG and video PPG datasets including our two custom datasets. Section IV describes the essential details of various neural network models for vitals estimation (including a vision transformer as well as CLIP caption generator model) that are trained on several datasets. Section V discusses the architecture of the proposed P2E-Net, which translates a video-PPG signal acquired from a smartphone into a single-lead ECG signal. Section VI concludes the paper.

\section{Related work}
The objective of this study is to extract a PPG signal from video-PPG data, which is obtained by recording a video while placing the index finger on the rear or front camera of a smartphone. To achieve this, we conduct a comprehensive review of existing research that estimates three body vitals--- pulse rate (PR), oxygen saturation (SpO2), and respiratory rate (RR)--- using both video-PPG signals and traditional PPG signals acquired via a pulse oximeter. Additionally, we provide a brief overview of the latest techniques for acquiring single-lead ECG using a smartphone.

\subsection{Video PPG-based Vitals Estimation}
Several studies have explored the use of video-PPG data to estimate body vitals, and a summary of these works is presented in Table I. However, the table highlights the limited availability of publicly accessible datasets for this purpose. The existing datasets either have a small number of examples or only provide labeled data for a subset of the vitals of interest\footnote{There are works under the name remote PPG or iPPG that measure the body vitals using dedicated cameras to record the face video and extract PPG signal from the video (this line of work was popular before the advent of smartphones), see \cite{mcduff} and references therein. Moreover, there is a recent flux of smart watches/smart wristbands (by Apple, Samsung, Fitbit, etc.) capable of measuring the body vitals. But since this work investigates the feasibility of smartphones as a physiological monitor, discussion of these works is out of the scope of this work.}. In contrast, this study aims to estimate not only the three vitals (pulse rate, oxygen saturation, and respiratory rate) but also the single-lead electrocardiogram (ECG), which is not addressed in the existing works



\begin{table}
\label{tab:lit_rev}
\caption{ Quick comparison of our work with the most relevant related works (all works utilize video-PPG data from the fingertip). PRV(HRV) stands for pulse (heart) rate variability, BP stands for blood pressure, HR stands for heart rate. }
\begin{center}
\begin{tabular}{|cp{2cm}|cp{2cm}|cp{2cm}|cp{2cm}|}
\hline

Ref.    & Vitals Measured & Dataset & Methodology \\\hline\hline
\cite{nemcova2020monitoring}    & PR, SpO2, BP   & private         &   Peaks detection   \\\hline
\cite{8579580}     & SpO2	& private           & SVD + CNN                 \\\hline
\cite{bui2020smartphone}     & SpO2	& private           &           --        \\\hline
\cite{nemcova2021brno}          & PR             & BUT PPG   &      --      \\\hline
\cite{neshitov2021wavelet}      & PR, PRV        & Welltory  & Wavelet Analysis     \\\hline
\cite{08989}      & HR, SpO2        & MTHS  & CNN \\\hline
\cite{pys_param} & HR, HRV, RR, SpO2   & --   & Peaks detection, VFCDM    \\\hline
This work & PR, SpO2, RR,  ECG (single-lead) & BIDMC and MTHS & CNN, DCT+FFNN      \\\hline 

\end{tabular}
\end{center}
\end{table}

{\it Pulse Rate:} 
The estimation of pulse rate has traditionally relied on specialized sensors that record PPG signals. However, later studies have explored the potential of mobile phones to record vPPG signals\footnote{{In this work, we use video-PPG and vPPG alternatively, to represent the PPG signal extracted from video acquired by a smartphone.}} and estimate vital signs. Early research, such as \cite{pys_param}, proposed using mobile phones to estimate heart rate (HR)\footnote{{In literature, HR and PR are used alternatively, therefore, we also use PR and HR alternatively in this paper.}}, and since then, smartphones have become widely studied as devices for measuring vital signs. Researchers have utilized various sensors available on mobile phones to monitor vitals. For instance, \cite{mohamed2017heartsense} proposed using a gyroscope to measure HR, while \cite{siddiqui2016pulse, zaman2017novel, nemcova2020monitoring} used the rear camera of the mobile phone to measure pulse rate using fingertip videos. The front camera has also been used to measure HR, as in \cite{kwon2012validation}. The vPPG signals have been recorded from various body locations, such as the face \cite{hassan2017heart, hernandez2020comparative}, fingertips \cite{siddiqui2016pulse, zaman2017novel, nemcova2020monitoring}, and forehead \cite{ruminski2016reliability} to extract vPPG for pulse rate estimation. Moreover, numerous methods have been proposed to measure vitals from vPPG signals, such as using Fourier transform and peak detection-based algorithms for pulse rate estimation in \cite{hoan2017real} and principal component analysis (PCA) in \cite{yu2015dynamic}. Some researchers have also employed deep learning (DL) models for pulse rate estimation, such as using a CNN in \cite{ayesha2021heart} and long short-term memory (LSTM) based attention network in \cite{gao2022remote}. Recently, researchers have explored the use of various transformer-based models for pulse rate estimation, such as vision transformer \cite{sun2023vit}, TransPPG \cite{kang2022transppg}, and Radiant \cite{gupta2023radiant}.


{\it Blood Oxygen Saturation (SpO2):} Since the smartphone cameras are not designed for pulse oximetry, smartphone-based SpO2 estimation faces several challenges, e.g., lack of infrared LED, lack of a well-accepted mathematical model, noisy vPPG signal, variable placement and pressure of fingertip on camera lens\cite{bui2017pho2}. Thus, authors in \cite{bui2017pho2} provide add-on hardware to measure SpO2 whose results meet the food and drug authority (FDA) accuracy standard when tested on six subjects only. Another smartphone-based pulse oximeter solution based on meta region of convergence (MROI), the ratio of ratio (RoR), and linear regression method are presented in \cite{9762399}. This method though that does not require any additional hardware has the limitation that it cannot estimate low SpO2 values. One of the earliest works utilizing the RoR method for mobile-based SpO2 estimation is \cite{scully2011physiological}; however, its accuracy is not up to the mark based on the FDA clearance threshold as it is also not able to estimate low SpO2 levels. For the patients of respiratory disease, a SpO2 estimation algorithm for low SpO2 level detection is presented in \cite{bui2020smartphone}. In \cite{kanva2014determination} RoR and linear regression method is used to measure SpO2. The RoR method uses green and red channel wavelength, the amplitude of the vPPG from different channels are different  and it depends on the quantum efficiency of the camera for different wavelengths. Thus, \cite{7145228} incorporated camera quantum efficiency in the measurement of SpO2 using a smartphone. Although the performance of this method is improved it suffers from a limitation of the unavailability of quantum efficiency of mobile cameras. To mitigate this problem, recently, authors in \cite{nemcova2020monitoring} proposed a method that neither requires information of quantum efficiency nor dedicated external hardware for the measurement of SpO2. Last but not least, the authors in \cite{ 8579580} estimated SpO2 using convolutional neural networks.

{\it Respiratory rate:}
Respiratory rate measurement has traditionally been carried out by manually counting chest movements, a time-consuming and inaccurate process \cite{philip2015accuracy}. Additionally, medical-grade equipment for RR measurement is costly and not widely available for use in wearable mobile devices \cite{charlton2016assessment}. However, mobile phones have become increasingly popular for measuring RR using various sensors. For instance, \cite{nam2015estimation} utilized the built-in microphone of a mobile phone to record nasal breath sounds for RR estimation. Similarly,  \cite{nam2014respiratory} measured RR using video recorded from the fingertip using a mobile phone's rear camera, utilizing three different methods: autoregressive model (AR), variable-frequency complex demodulation (VFCDM), and continuous wavelet transform (CWT). Aly et al. \cite{aly2016zephyr} utilized the accelerometer and gyroscope of a mobile phone held on a human chest to extract RR, while  \cite{alafeef2020smartphone} used the discrete wavelet transform to measure RR from video recorded from the fingertip, extracting the vPPG. Moreover, several studies have investigated deep learning (DL)-based approaches for RR estimation, with Shuzan et al. \cite{shuzan2023machine} recently investigating 19 DL models for estimating RR and HR, with the Gaussian process regression model demonstrating the best performance.


\begin{figure*}
\centering

    \includegraphics[width=0.9\textwidth]{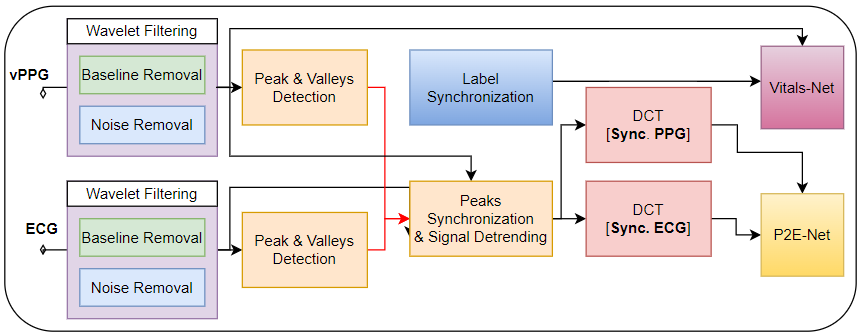}
    \caption{{Highlights of the pre-processing done for vitals estimation and vPPG to ECG reconstruction.}}
    \label{fig:preprop}
\end{figure*}

\begin{figure*}
\centering

\begin{subfigure}{0.32\textwidth}
    \includegraphics[width=1\textwidth]{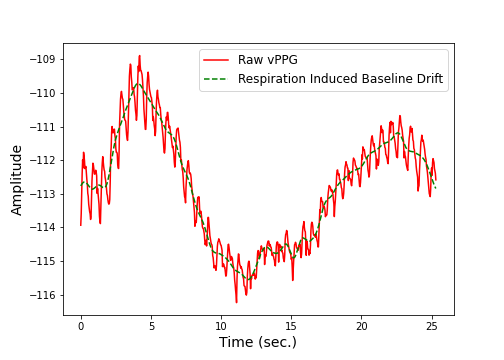}
    \caption{PPG signal with respiration induced baseline and noise.}
    \label{fig2:a}
\end{subfigure}
\hspace{0.001\textwidth}
\centering
\begin{subfigure}{0.327\textwidth}
    \includegraphics[width=1\textwidth]{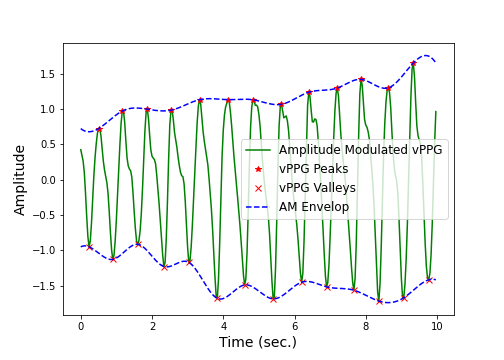}
    \caption{ PPG signal with respiration induced baseline (but with noise removed).}
    \label{fig2:b}
\end{subfigure}
\centering
\hspace{0.001\textwidth}
\begin{subfigure}{0.32\textwidth}
    \includegraphics[width=1\textwidth]{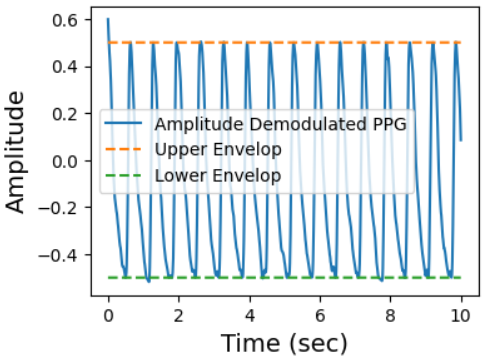}
    \caption{PPG signal after removal of respiration induced amplitude modulation and noise.}
    \label{fig2:c}
    
\end{subfigure}
\hspace{0.5\textwidth}
\caption{{Detrending and denoising: two key steps in the preprocessing of a PPG signal.}}
\label{fig:2}
\end{figure*}

\subsection{Single-lead ECG Reconstruction} 
There are only a couple of works that have considered generation of a single-lead ECG signal using a smartphone and/or a pulse-oximeter. For example,  \cite{sagirova2021cuffless} mounts two dry electrodes on the back-casing of the smartphone in order to measure the single-lead ECG by placing one's both thumbs on the two dry electrodes. \cite{zhu2019ecg} passes the PPG signal acquired from a pulse oximeter through a block that computes the discrete cosine transform, followed by ridge regression, in order to get the single-lead ECG signal. \cite{vo2021p2e} also attempts to translate a PPG signal (acquired by a pulse oximeter) to a single-lead ECG signal, but by using a conditional generative adversarial network (c-GAN).  


\subsection{Research Gap} 
To the best of the authors' knowledge, a stand-alone smartphone-based solution that measures the body vitals (PR, SpO2, RR), as well as a single-lead ECG (without any external add-on hardware/sensors), has not been discussed in the open literature to date. 
Moreover, existing video-PPG-based frameworks are optimized for some specific scenarios on small datasets. Work on the generalization capability and robustness of the DL-based vitals estimation model is scarce. 



\subsection{Contributions}
Having motivated the research gap, the main contribution of this paper is two-fold:
\begin{itemize}
\item {\it Estimation of body vitals:} We pre-process the video-PPG signal acquired from a smartphone and feed it to various custom-built DL models (including a vision transformer and a CLIP model) which eventually output the three most important body vitals (PR, SpO2, RR). We do hope that our dataset will serve as a benchmark dataset to test the generalization capability, accuracy and robustness of future deep learning methods for vitals estimation. 
\item {\it Synthesis of single-lead ECG:} We pre-process the video-PPG signal acquired from a smartphone and utilize a novel discrete cosine transform+feedforward neural network-based method that translates the recorded video-PPG signal to a single-lead ECG signal. To the best of our knowledge, this is the first work that reconstructs a single-lead ECG from a video-PPG signal acquired from a smartphone.
\end{itemize}

\section{Datasets}

{In this subsection, we outline the details of our K20-vPPG and K1-vP2E datasets. But before that, it is imperative to provide the reader with a systematic and brief review of the relevant existing public datasets on PPG and vPPG (as well as their limitations)\footnote{Note that there are PPG datasets constructed using commercial pulse oximeters, e.g., MIMIC-III dataset \cite{johnson2016mimic}, \cite{liang2018new}, video-PPG datasets based upon traditional cameras \cite{mcduff} (before the advent of smartphones), and video-PPG datasets that utilize PPG signals for biometrics/authentication purposes, e.g., biosec1 dataset \cite{8411233}. But since this work primarily focuses on video-PPG datasets acquired using smartphones for vitals estimation, discussion of such datasets (and the corresponding works) are out of the scope of this work.}.}

\subsection{Existing video-PPG datasets}

\textit{1) BUT-PPG dataset \cite{nemcova2021brno}}:   
This dataset contains 48 simultaneous records of video-PPG data and single-lead ECG data of 12 healthy subjects (6 males, 6 females), 21 to 61 years old. Each video of the index finger of the subject was recorded for a duration of 10 seconds, using a Xiaomi MI9 smartphone at a frame rate of 30Hz. The single-lead ECG data were recorded using a Bittium Faros 360 device at a sampling rate of 1000 Hz and were manually annotated by an expert. Eventually, \cite{nemcova2021brno} utilized their dataset to estimate the HR.

\textit{2) Welltory dataset \cite{neshitov2021wavelet}}:  
This dataset contains 21 records of video-PPG data of 13 healthy subjects, 25 to 35 years old. Each video of the index finger of the subject was recorded for a duration between 1-2 minutes, using Welltory android app. The R-peak to R-peak (RR) intervals were recorded using a Polar H10 ECG chest strap and were manually examined by an expert. Eventually, \cite{neshitov2021wavelet} utilized their dataset to estimate the HR and HRV.

\textit{3) MTHS dataset \cite{08989}}:  
This dataset contains 65 recordings of video-PPG data along with corresponding HR and SpO2 labels, of 62 patients (35 males, 27 females). Each video of the index finger of the subject was recorded using an iPhone 5s smartphone at a frame rate of 30 Hz. The ground truth/labels were obtained using a pulse oximeter (M70) at a sampling rate of 1 Hz. Eventually, \cite{08989} utilized their dataset to estimate the HR and SpO2.

{\it Limitations of existing vPPG datasets:}
 We identified some limitations/shortcomings of the aforementioned vPPG datasets, some of them are as follows. The BUT-PPG dataset provides labels for HR only, while the Welltory dataset provides labels for HR and HRV only. Moreover, the small size of BUT-PPG and Welltory datasets renders them infeasible for state-of-the-art but data-hungry deep learning methods. The MTHS dataset though contains sufficient examples (enough to train a neural network) but provides labels for HR and SpO2 only. More importantly, the need for a new and large dataset on video-PPG stems from the fact that the generalization capability, accuracy, and robustness of any deep learning algorithm for vitals estimation could only be tested if a handful of video-PPG datasets are publicly available.

\subsection{Benchmark PPG datasets}

{1) The BIDMC dataset:}
As we mentioned earlier in this section, one could learn more about the generalization capability, accuracy, and robustness of his/her proposed deep learning algorithm by testing it on other datasets (with unseen data with potentially different distributions). Therefore, this work utilizes the well-known BIDMC dataset \cite{pimentel2016toward} (in addition to the K20-vPPG dataset) for the training and performance evaluation of the proposed method for vitals estimation. Some most pertinent details of the BIDMC dataset are as follows. The BIDMC dataset contains 53 sessions (each of duration eight minutes) of simultaneously recorded PPG and ECG signals, along with the ground truth values (i.e., the vitals). The PPG and ECG signals are recorded at a sampling frequency of 125 Hz, whereas the ground truth values of HR, SpO2, and RR are recorded at a sampling rate of 1 Hz. Note that the PPG signals in this dataset were acquired from the fingertip of patients using the clinical pulse oximeter. Finally, the single-lead ECG signal collected in this dataset is the Lead-II acquired using the standard 12-lead ECG.  

{2) The PulseDB dataset\cite{wang2022pulsedb}:}
{This dataset contains a large number of filtered PPG and ECG signals. It also contains the ground truth labels for HR and BP. We randomly download the data of 550 subjects both male and female with more than 16000 PPG, and ECG signals along with their corresponding vitals ground truth labels. Each PPG and ECG signal is 10 seconds long and sampled at 125 Hz. The corresponding ground truth labels are recorded at 1 Hz.}

\subsection{Our video-PPG datasets}
{
{\it A) K20-vPPG dataset:}
The limitations of the existing vPPG datasets (e.g., a small number of training examples, lack of raw data and labels for RR, and single-lead ECG, as needed by our study) prompted us to run an extensive campaign for vPPG data collection of our own. Thus, we subsequently compiled a new dataset named K20-vPPG dataset. The data collection campaign was approved by the ethical institutional review board (EIRB) of our institution, and all the subjects voluntarily participated in this data collection activity. Next, we discuss all the relevant details of the K20-vPPG dataset.
}

{\it Participants:}
{
A total of 20 healthy subjects with no history of cardiac or respiratory disease participated in this data collection campaign, of which 5 were females and 15 were males. The volunteers/participants were either employees or students at our institute, aged 16-36 years. 
}

{\it Data characteristics:} 
{
For each subject, we recorded the 2 to 10 minutes long vPPG data (the raw data) from the index fingers of the right hands of twenty different subjects. For ground truth/labels for supervised learning later, we simultaneously recorded the three body vitals (PR, SpO2, RR)\footnote{ {This dataset also contains the ground truth labels of perfusion index (Pi) and Pleth Variability Index (PVi). However, their discussion is out of the scope of this paper.}} 
}

{\it B) K1-vP2E dataset:} 
{For training the P2E-Net, the lead author simultaneously recorded 24, 5-10 minutes long vPPG and ECG signals of himself over a time period of seven days after different activities (e.g. eating, running, sleeping, and walking). Then, the signals were filtered and detrended in a similar way shown in Fig.} \ref{fig:2}.

\section{Vitals Estimation using video-PPG}

The main objective of this section is to conduct a comprehensive evaluation of DL-based models for vitals estimation, focusing on their generalization capability, accuracy, and robustness. Previous studies have highlighted that existing video-PPG and DL-based approaches are extensively parameterized and optimized on small public datasets. While these models may exhibit strong performance on these datasets, they often lack the ability to generalize and demonstrate robustness in diverse scenarios. Thus, our contribution lies in enhancing the generalization capability and robustness of various DL-based architectures. The section begins by discussing crucial pre-processing steps, followed by a detailed description of our proposed DL-based model, Vitals-Net. Finally, we conclude with a thorough performance evaluation of our method.


\begin{figure*}
     \centering
     \begin{subfigure}[b]{0.45\textwidth}
         \centering
         \includegraphics[width=1.1\textwidth]{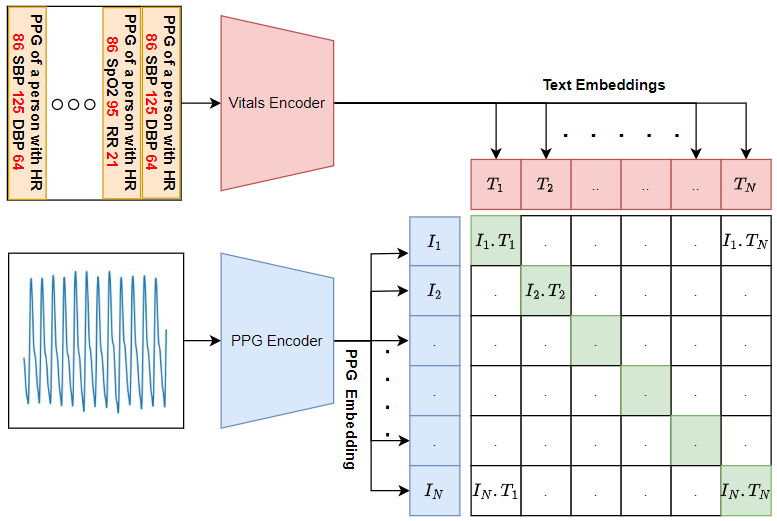}
         \caption{Training Vitals-CLIP model}
         \label{fig:CLIP_Train}
     \end{subfigure}
     \hfill
     \begin{subfigure}[b]{0.45\textwidth}
         \centering
         \includegraphics[width=1.1\textwidth]{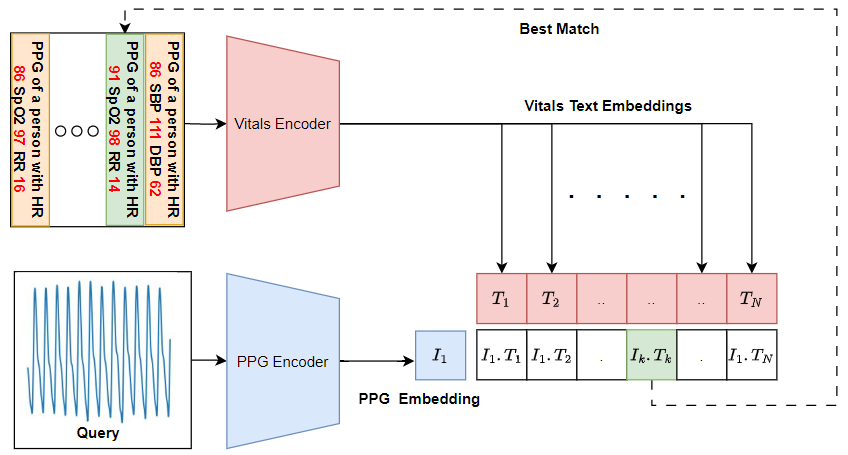}
         \caption{Vitals Estimation using Vitals-CLIP.}
         \label{fig:CLIP_test}
     \end{subfigure}
     \hfill
             \caption{The CLIP Neural Network model for one-shot estimation of Vitals.}
        \label{fig:Vitals_results}
\end{figure*}

\begin{figure}

    \includegraphics[width=0.5\textwidth]{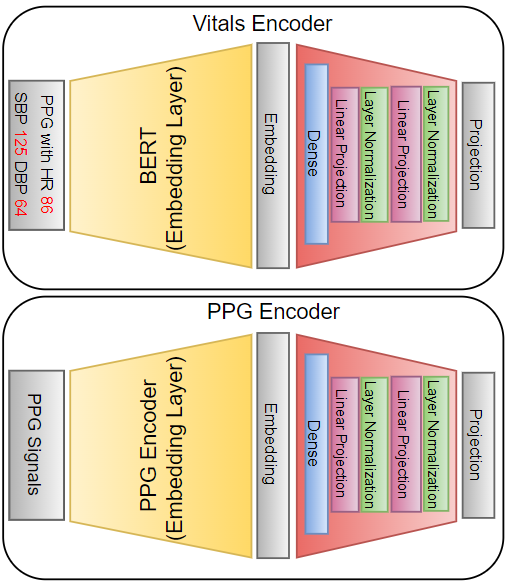}
    \caption{The model architecture details of Encoders in Vitals-CLIP. The inner architecture of the PPG Encoder's embedding layer and linear projection layer is given in Table \ref{tab:model_architect}}
    
\label{fig:encoders}
\end{figure}

\subsection{Pre-Processing Stage}
\label{pre-proc}

The pre-processing stage for video-PPG data, including the PPG signal, is performed prior to training the vitals-Net model. Careful pre-processing not only enhances the training accuracy of the neural network but also facilitates the model's training process. Our pre-processing stage has been specifically designed to address various distortions, such as baseline drift, high-frequency noise, artifacts caused by ambient light and motion, and the de-trending of the PPG signal. The following are the detailed pre-processing steps (and are summarized in Fig. \ref{fig:preprop}):

\textit{PPG signal extraction from video-PPG:} The pixel averaging technique is employed to extract a PPG time series from the central region of each frame in the video-PPG. This yields a vPPG signal, which is combined with simultaneously recorded ground truth vital signs obtained from an oximeter for training and validation purposes.

\textit{Sliding window mechanism}: Since the ground truth labels for vitals are recorded at a frequency of 1 Hz, a sliding window mechanism is utilized. This mechanism uses a stride of 1 second, allowing the creation of a video PPG dataset. By employing this mechanism, we extract maximum PPG and corresponding vital sign examples in the dataset.


    

\textit{Wavelet Filtering--} In this filtering stage, the vPPG signal undergoes wavelet filtering to eliminate low-frequency baseline drift, high-frequency noise, artifacts induced by ambient light, and motion artifacts. A five-level decomposition is applied to the vPPG signal to obtain approximate and detailed coefficients. Subsequently, the signal is reconstructed by selectively choosing appropriate wavelet coefficients, as described in \cite{ahmed2023deep}. The low-frequency baseline induced by respiration and changes in ambient light intensity is represented by the signal corresponding to the approximate coefficients. The high-frequency noise and motion artifacts are represented by the signals corresponding to the level 1 and 2 detailed coefficients, respectively. Therefore, the reconstructed signal consists of wavelet coefficients other than the approximate and detailed coefficients from the first two levels of decomposition. Fig. \ref{fig2:a} illustrates the input raw PPG signal, while Fig. \ref{fig2:b} shows the corresponding filtered signal but with baseline intact. Finally, Fig. \ref{fig2:c} showcases the de-trended and de-noised PPG signal, which is particularly useful for ECG reconstruction.


\subsection{Vitals-Net models and training}
\subsubsection{Vitals-Net models}
We train four different models namely CNN-Net, MT-Net, ViT-Net, and ResNet, to estimate the vitals. Additionally, we train the Vitals-CLIP model for querying-based vitals estimation. 

\textit{CNN-Net model:} This model is a variant of a CNN model proposed in \cite{reiss2019deep} which is originally proposed for heart rate estimation; however, we fine-tuned this model to estimate all vitals. Specifically, we add a lambda layer, that computes a short-time Fourier transform of the input PPG signal, at the top of this model. Moreover, we add batch normalization followed by a dropout layer after the flattening layer and each dense layer to avoid over-fitting and ease the training process.
We trained four different models: CNN-Net, MT-Net, ViT-Net, and ResNet, for vital sign estimation. Additionally, we trained a Vitals-CLIP model for querying mode vitals estimation.

\textit{MT-Net model:} This model is another fine-tuned model originally proposed for HR and SpO2 estimation using vPPG \cite{samavati2022efficient}.

\textit{ViT-Net model:} This model is a fine-tuned vision transformer. For compatibility, we added a lambda layer capable of performing a short-time Fourier transform of the input PPG signal. It is worth noting that the ViT model inherently expects a 2D vector/matrix as input, and the short-time Fourier transform converts the PPG signal into a suitable 2D vector/matrix format, making the lambda layer appropriate for integration with the ViT model.

{\it Vitals-CLIP model architecture:}
\
Vitals-CLIP is an enhanced version of the audioCLIP model \cite{guzhov2022audioclip}, which extends its capabilities by incorporating PPG signals along with the text. To achieve this enhancement, we integrate specialized encoder models for PPG and text encoding, which consists of a text embedding layer followed by a projection layer (see Fig. \ref{fig:encoders} top) and PPG embedding layer followed by a projection layer (see Fig. \ref{fig:encoders} bottom),  into the existing CLIP framework, leveraging the vPPG and PulseDB dataset. This integration allows our proposed model to perform both bimodal and unimodal querying tasks while maintaining CLIP's impressive generalization capability to novel datasets.

The model comprises three encoder models: Vitals Encoder, PPG Encoder, and ECG Encoder. The Vitals Encoder consists of a text encoder, that generates 
embeddings of the text input, followed by an embedding projection model. We utilize a pre-trained BERT model obtained from TensorFlow Hub. The embedding projection layer in the Vitals Encoder includes a Dense layer, followed by two linear projection layers. Each linear projection layer consists of a 'gelu' activation layer, a dense, a dropout layer with a dropout rate of 0.2, an ADD layer that adds the output of the dropout layer and the first dense layer outside the linear projection layer, and a 'layernormalization' layer at the end\footnote{{Specifically, the pre-trained BERT model that we have used could be downloaded from the following URL: https://tfhub.dev/tensorflow/small\_bert/bert\_en\_uncased\_L-4\_H-512\_A-8/1}}. 

\begin{table}

    \caption{ {Model architecture of PPG/ECG Encoders in CLIP. Here, $N_f$,$K_s$, and $s$ denote the number of filters, kernel size, and stride respectively. }}
\begin{center}
\begin{tabular}{|c|*{6}{c|}}
\hline
& \multicolumn{3}{c|}{PPG/ECG Embedding }  & \multicolumn{2}{|c|}{Linear Projection}  \\\hline\hline
Layer& Type & Output & ($N_f$,$K_s$,$s$) &Layer & Output  \\\hline\hline

1    & Input  & (1250,1)     & -         &     GELU     & (256)   \\\hline
2    & Conv1D & (1241,8)     & (64,10,1) &     Dense	& (256)   \\\hline
3    & MaxPooling1D	& (620,8)& -         &     Dropout  & (256)   \\\hline
4    & Conv1D  & (616,8)     & (32,5,1)  &     ADD      & (256)   \\\hline
5    & MaxPooling1D & (308,8)& -         &           &    \\\hline
6    & Conv1D  & (306,16)    & (16,3,1)  &              &          \\\hline
7    & Conv1D  & (304,32)    & (8,3,1)   &              &          \\\hline
8    & Flatten & (9728)      & -         &              &          \\\hline 

\end{tabular}
\end{center}
\label{tab:model_architect}
\end{table}

The PPG Encoder and ECG Encoder have same base architecture. The PPG/ECG encoder consists a 1D CNN based model, that output low dimentional embeddings of both PPG and ECG signal, followed by same embedding projection layer described above. More specifically the architecture of PPG/ECG encoder is summarized in the Table \ref{tab:model_architect}.

\subsubsection{Training}
We train our models in three different configurations
\textit{Vital Specific Training Configuration--} From the literature survey, it is clear that training and testing of DL models are performed independently for each vital. Therefore, we train our models first in a vital specific configuration. In this configuration, we train our models for each individual vitals. In this configuration, we train our model using a video PPG dataset and then by BIDMC dataset. The video PPG dataset contains $21-30$, $5-10$ minutes long vPPG signals that were obtained by averaging each frame of fingertip video. Similarly, the BIDMC contains $8$ minutes long each from $53$ subjects. To train these models we segment each signal in BIDMC and K20-vPPG datasets in a vital specific window size. For example, we segment each PPG using $w_s = 10\ sec$ for vitals HR and RR and $w_s = 30\ sec$ for RR. For each segment, we use an average of 10 labels for HR/SpO2 and an average of 30 RR labels for RR estimation. By the segmentation, we get a dataset $D \in \mathcal{R}^{N_v, N \times w_s*Fs \times ch }$ where $N$, $N_v$ and $ch$ denotes the number of PPG and label pairs, number of vitals and the number of channels, e.g. $ch = 2$ means red and green video PPG signals, respectively. The dimensions of  BIDMC and vPPG datasets are  $\mathcal{D}_b = \mathcal{R}^{\{N_v, 22550\times30*w_s \times 1\}}$ and $\mathcal{D}_v = \mathcal{R}^{\{N_v, 8890\times30*w_s\times 2\}}$  respectively. Other training parameters are added in Table \ref{tab:table2}. It is well known that PPG datasets are prone to outliers due to the estimation error of oximeters, therefore these models must be trained in such a way that they train robustly on datasets that contains some outliers. Fortunately, mean absolute errors (MAE) are known to be robust to outliers. Therefore, we train our models using the MAE loss function defined below,
\begin{equation}
    \mathcal{L} = \frac{1}{|\mathcal{B}| }\sum_{\mathbf{y} \in \mathcal{B}} |\mathbf{y}-\mathbf{\hat{y}}|
\end{equation}
where $\mathcal{B}$ is the batch size, $\mathbf{y}$ is the label, $\mathbf{\hat{y}}$ is the prediction.
With the above loss function, we train the model for a maximum of 1000 epochs, however, we apply an early stopping to avoid overfitting of models.

{\textit{Joint Training Configuration}--}  In many scenarios where these vitals estimation is performed using low resources devices e.g., wristbands, smart watches, or even mobile, estimation of each vital with different DL models is not resource efficient. Therefore, in contrast to the first configuration where we trained models for only one vital, in this configuration we stack all vitals in a vector  and use that as a label while training the model. Similar to the previous configuration, in this configuration DL models are trained on both BIDMC and vPPG datasets independently. In contrast to the previous configuration, the $w_s$ is the same for all vitals. In this configuration, we use $w_s = 20\ sec$.
The number of PPG and labels paired in the BIDMC and vPPG datasets are the same as the previous configuration.

{\textit {CLIP Configuration}--} In contrast to the previous two configurations, this configuration is specifically related to the Vitals-CLIP model and differs from the previous configurations in a number of ways. Firstly, this configuration involves a so-called pretraining which is totally different from the training first two configurations. Secondly, this configuration uses text captions (see Fig. \ref{fig:caption} and Fig. \ref{fig:CLIP_Train}) as labels rather than numeric labels. Thirdly, the pretraining is not dataset specific rather three datasets namely vPPG, BIDMC, and PulseDB, are used simultaneously in the pertaining. This is possible due to the fact that text captions can be of different sizes. We train Vitals-CLIP in this configuration on a very large dataset. To make the dataset larger we made use of the subset of the PulseDB dataset in addition to BIDMC and video PPG. We used a large number of $10$ seconds long PPG signals from 550 subjects taken randomly from the PulseDB dataset and concatenate it with the BIDMC dataset containing the PPG of 53 subjects and also concatenated the video PPG dataset with them. Using the  ground truth values of vitals, we made captions corresponding to each PPG segment. 
Finally, we get a dataset $\mathcal{D}_C \in \mathcal{R}^{\{N\times N_c,N \times w_s*F_s \times 1 \}}$ where $N = 38600$, $N_c$ is caption length, $F_s = 125$ Hz and $w_s = 20$ seconds.

We use this dataset for the training of Vitals-CLIP. After the pretraining of Vitals-CLIP, we use it in querying mode (see Fig. \ref{fig:CLIP_test}) due to its inherent ability to operate in querying mode. After pretraining we generate text embedding all labels in the validation dataset and search for a caption using a ppg signal as a query. The caption searching is performed by generating an embedding using a pre-trained PPG encoder and then taking the dot product with all caption/text embeddings. Then, the corresponding caption of the top $k$ dot products are selected as captions (see Fig. \ref{fig:caption}). Then from  all $k$ captions the values of vitals are extracted with Python function and taken as an average of them and then compare with ground truths.

\begin{figure}
     \centering
     \begin{subfigure}[b]{0.24\textwidth}
         \centering
         \includegraphics[width=1.0\textwidth]{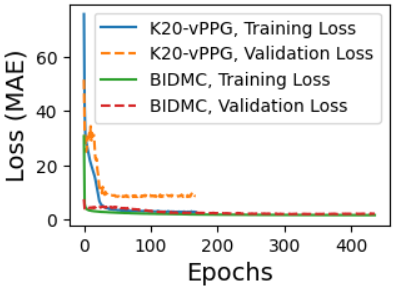}
         \caption{}
         \label{fig:HR_loss}
     \end{subfigure}
     \hfill
     \begin{subfigure}[b]{0.24\textwidth}
         \centering
         \includegraphics[width=1.0\textwidth]{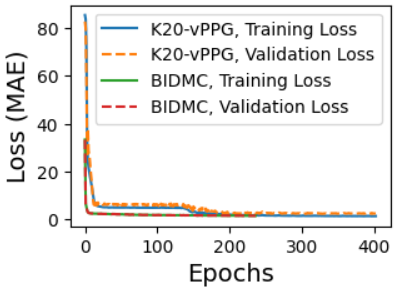}
         \caption{}
         \label{fig:oxi_loss}
     \end{subfigure}
     \hfill
     \begin{subfigure}[b]{0.235\textwidth}
         \centering
         \includegraphics[width=1.0\textwidth]{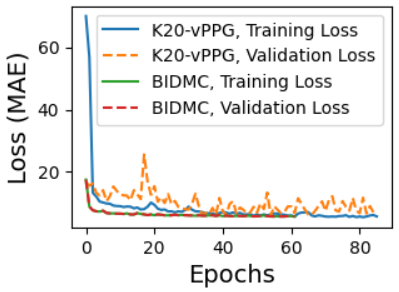}
         \caption{}
         \label{fig:rr_loss}
     \end{subfigure}
      \hfill
     \begin{subfigure}[b]{0.235\textwidth}
         \centering
         \includegraphics[width=1.0\textwidth]{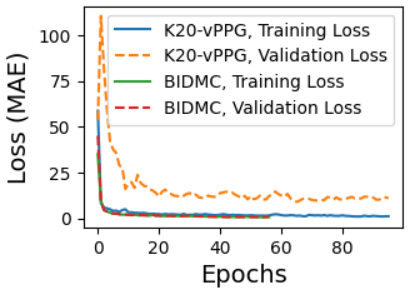}
         \caption{}
         \label{fig:hr_error}
     \end{subfigure}
     \hfill
    \caption{Training and validation
loss of fine-tunned (a)  CNN-Net (b) MT-Net (c) ViT-Net (d) ResNet50 models using vPPG and
BIDMC datasets.}
\label{fig:losses}
\end{figure}

\begin{figure*}
     \centering

     \hfill
     \begin{subfigure}[b]{0.32\textwidth}
         \centering
         \includegraphics[width=1.0\textwidth]{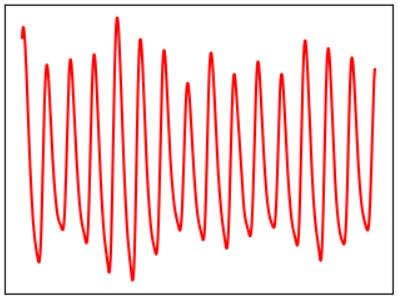}
         \caption{GT: A PPG with HR 91 SBP 59 DBP 42.\\ 
         C1: A PPG with HR 94 SpO2 99 RR 17. \\ 
         C2: A PPG with HR 94  SBP 58 DBP 44. \\
         C3: A PPG with HR 91  SBP 61 DBP 42. \\
         }
         \label{fig:oxi_loss}
     \end{subfigure}
     \hfill
     \begin{subfigure}[b]{0.32\textwidth}
         \centering
         \includegraphics[width=1.0\textwidth]{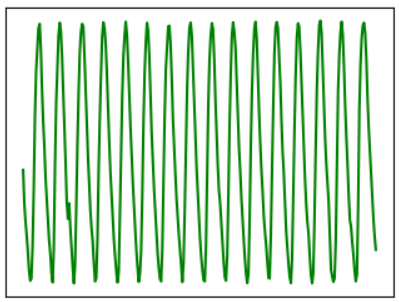}
         \caption{GT: A PPG with HR 98 SpO2 98 RR 16.\\ 
         C1: A PPG with HR 98 SpO2 99 RR 09. \\ 
         C2: A PPG with HR 100 SpO2 58 RR 09. \\
         C3: A PPG with HR 96 SpO2 61 RR 13. \\
         }
         \label{fig:rr_loss}
     \end{subfigure}
      \hfill
     \begin{subfigure}[b]{0.32\textwidth}
         \centering
         \includegraphics[width=1.0\textwidth]{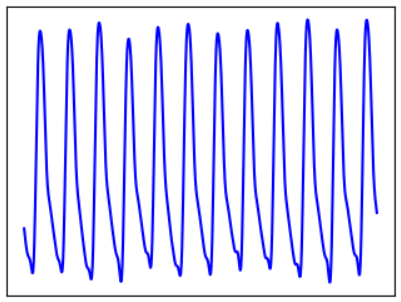}
         \caption{GT: A PPG with HR 73 SpO2 98 RR 19.\\ 
         C1: A PPG with HR 76 SpO2 99 RR 17. \\ 
         C2: A PPG with HR 78 SpO2 98 RR 14. \\
         C3: A PPG with HR 82 SBP 88 DBP 52. \\
         }
         \label{fig:hr_error}
     \end{subfigure}
     \hfill
    \caption{Top three matching captions in CLIP for the PPG signals from (a) PulseDB dataset (b) BIDMC dataset (c) K20-vPPG dataset. GT stands for ground truth. }
    \label{fig:caption}
\end{figure*}
\begin{table*}

\caption{\label{tab:table3} Absolute error performance of the DL Architecture proposed in this work on the two datasets (K20-vPPG and BIDMC).}

\begin{center}
\begin{tabular}{|c||*{12}{c|}}
\hline

\multicolumn{13}{|c|}{Vital Specific Training Configuration}     \\\hline
~ &\multicolumn{6}{c|}{K20-vPPG}   &   \multicolumn{6}{c|}{BIDMC}  \\\hline

~ &\multicolumn{2}{c|}{HR}  & \multicolumn{2}{c|}{SpO2}  &\multicolumn{2}{c|}{RR}   & \multicolumn{2}{c|}{HR}  & \multicolumn{2}{c|}{SpO2}  &\multicolumn{2}{c|}{RR}  \\\hline

DL Models & $\mu$ & $\sigma$ & $\mu$ & $\sigma$  & $\mu$ & $\sigma$ & $\mu$ &  $\sigma$  & $\mu$ & $\sigma$ & $\mu$ & $\sigma$  \\\hline


CNN-Net & 2.22 & 4.04 & 0.64 &  0.61 & 3.95  & 2.78	&
4.10 & 2.91 & 5.91 & 5.21 & 2.85 & 2.24 \\\hline
MT-Net & 2.07& 4.00 & 1.99 &  1.53 & 3.58  & 2.45	&
4.19 & 2.48 & 6.12 & 7.33 & 2.67 & 2.67 \\\hline
ViT-Net  & 4.11& 3.74 & 2.22& 2.01 & 4.31  & 3.36 & 4.36 &  4.3 &3.01  &6.59  & 2.44& \\\hline
ResNet50 &3.32   &3.83 & 2.26 & 1.91  & 2.62  & 3.57	&
4.20 & 16.56  & 6.66 & 6.14& 4.99 & 3.82 \\\hline
\multicolumn{13}{|c|}{Joint Training Configuration}     \\\hline

CNN-Net & $5.99$  & $6.28$  & $1.05$  & $1.16$&$3.46$ & $2.92$ & 4.78 & 16.63 & 93.75 & $17.57$ & $3.10$ & $2.80$  \\\hline

MT-Net & 5.62 & 6.22& 0.64 &  0.61 & 3.16  & 2.72	&
4.64 & 16.97& 5.44 & 16.85 & 2.54 & 1.98 \\\hline
ViT-Net  & 5.88 & 5.00 & 2.35 &  2.01 & 3.20  & 2.7 & 5.88 & 5.00  & 2.35 & 2.01 &3.20& 2.70\\\hline
ResNet50 & 6.17 & 7.92& 1.61 &  0.54 & 2.62  & 3.57	&
4.20 & 16.56  & 6.38 & 6.76 & 4.99 & 3.82 \\\hline
\multicolumn{13}{|c|}{CLIP Configuration}     \\\hline
Vitals-CLIP & 5.91 & 4.22 & 2.01 &  4.03 & 3.11  & 3.57	&
4.43 & 4.01  & 2.31 & 2.02 & 3.63 & 4.01 \\\hline

\end{tabular}
\end{center}
\end{table*}



\subsection{Performance Evaluation}
\begin{figure}
\begin{subfigure}{0.5\textwidth}
    \includegraphics[width=0.85\textwidth]{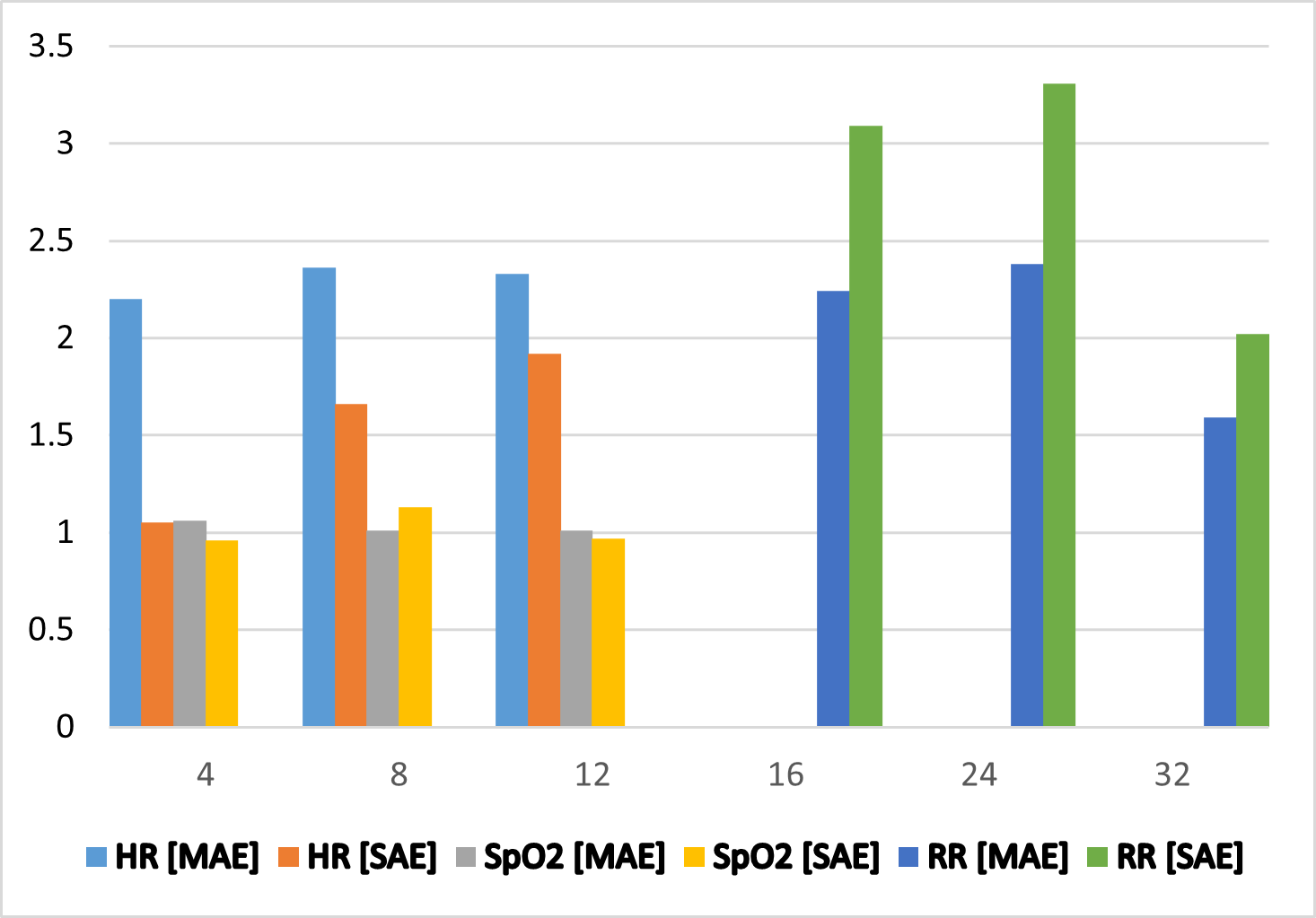}
    \label{fig:Ws_impct}
\end{subfigure}

\caption{Impact of Window Size on Vitals Estimation (x-axis represents window size, y-axis represents MAE and SAE).}
\label{fig:opt_chanel}
\end{figure}

To see the generalization capability of the DL model, we train these models with K20-vPPG and BIDMC datasets and then test them on the LESSO data of both datasets. The performance evaluation using LESSO data of video-PPG better indicates the generalization capability of a model\cite{reiss2019deep}. Therefore, we left some subjects out (LESSO) to use for the testing of the trained model. Specifically, we left data of some 5 subjects randomly selected from each dataset that serves as LESSO data. First of all, we use the vPPG dataset and figure out the best vPPG signal length denoted by $w_s$ (window size) for which the MAE is the minimum for each vital.  we perform further simulations in search of optimal video-PPG channel and window size. Fig. \ref{fig:opt_chanel} shows that the MAE for parameters HR and SpO2 increase slightly with the increase in widow size because the ground truth labels are the average of all labels in that window size. Overall, the window size of 4 seconds  proves to be the best window size for HR and SpO2 Estimation. The standard deviation in absolute error fluctuates slightly but remains under $1.5$. In contrast, we use a higher window size due to the fact that RR induces slow variations in the video-PPG signal. Both the MAE and SAE increase slightly and then decrease reaching their lowest values at the window size of 32 seconds.

After finding the best window size, we train these models separately for each vital, and their training and validation losses are not added here due to space constraints. Then, we train four models jointly using all vitals, and Fig. \ref{fig:losses} shows their training and validation losses when trained using two datasets vPPG and BIDMC. More specifically figure shows that the training and validation losses saturate after $200$ epochs in case of CNN-Net and MT-Net. However, the training and validation losses saturate earlier closer to 20 epochs for ViT-Net and ResNet50 models. The detailed results of all these models, trained jointly or vital specific configuration, are shown in Table \ref{tab:table3}. Overall the performance of models trained specifically for one vital is superior to the models trained using all vitals simultaneously. Overall MT-Net outperforms all these models. We also provide a preliminary study on the use of Vitals-CLIP in the querying mode to estimate the vitals. Table \ref{tab:table3} shows the detailed performance of Vitals-CLIP. 
It shows that Vitals-CLIP outperforms CNN-Net and ResNet50 in the estimation of HR when trained in joint training configuration. Also, Vitals-CLIP outperforms all models in SpO2 estimation for the BIDMC dataset. Also, here it is worth noting that almost all models fail to estimate SpO2 using a single PPG signal of the BIDMC dataset. Note that SpO2 estimation requires two PPG signals recorded using two lights of different wavelengths. The reason behind the failure of most of the models to predict SpO2 is that BIDMC dataset only has one PPG signal available in contrast to the video PPG where three PPG signals are available  e.g. from RGB channels. Here it is worth noting that Vitals-CLIP outperforms other methods in SpO2 estimation due to the fact that Vitals-CLIP works in querying mode. 

Fig. \ref{fig:caption} shows three different query-PPG signals from three different datasets and their corresponding top three matching captions and the ground truth caption. After getting the matching caption the value of labels extracted from caption strings and then taken average. From the figure it is worth noting that the matching caption can sometimes also gives the values of systolic blood pressure (SBP) and diastolic blood pressure (DBP). This is due to the fact that the PulseDB dataset contains three two vitals HR and blood pressure (BP) and some time query PPG signal matches with a caption of the PulseDB dataset. Here it is worth mentioning that due to the unavailability of BP labels for the vPPG and BIDMC datasets, we could not add SBP and DBP vitals estimation performance in the table. However, it can be deducted from the nature of captions that the performance of SBP and DBP estimation would be similar to the other vitals.

\begin{figure}
\centering
    \includegraphics[width=0.15\textwidth]{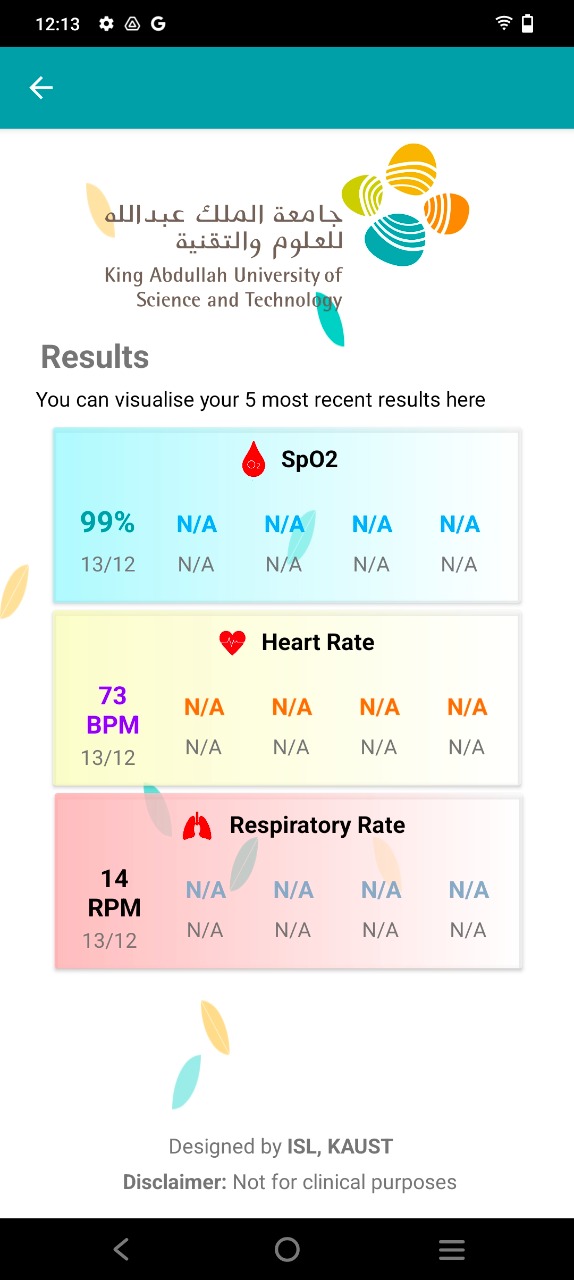}
    \caption{Screenshot of a custom android app developed using the proposed Vitals-net framework.}
    \label{fig:app}
\end{figure}

After the customization of the DL model of Vitals-Net \footnote{In this paper, Vitals-Net collectively refers to CNN-Net, MT-Net, ViT-Net, and ResNet50.}, we convert them into mobile-compatible models using the TensorFlow lite package.  The computational complexity of the converted model reduces further. We, then, make a customized app for vitals estimation using these Vitals-Net converted to TensorFlow lite model. Fig. \ref{fig:app} shows a screenshot of the custom-designed Android app that shows how to record a video and eventually displays the results after pre-processing and deep learning. 

\section{Single-lead ECG synthesis from video-PPG}

This section aims to reconstruct/synthesize a single-lead ECG signal from a PPG signal which itself has been extracted from the video-PPG data that is acquired by placing the fingertip on the rear camera of a smartphone\footnote{More precisely, we aim to reconstruct the ECG Lead-I, as per nomenclature for a standard 12-lead ECG system. }. Mathematically, the problem at hand is to find a mapping from a function $x(t)$ to another function $y(t)$, and vice versa. This (translation or regression) problem is indeed feasible due to the fact that the two signals (PPG and ECG) are highly correlated as they both capture the same cardiac activity at a sub-cardiac cycle resolution. Further, the morphology of the two signals is tightly binded to each other, from one cardiac cycle to another (e.g., the R-peak of the ECG corresponds to the diastolic peak of the PPG signal, and more). Next, as we did in the previous section, we first describe the crucial pre-processing steps, followed by the details of our proposed deep learning-based model (P2E-Net), followed by the performance evaluation of the proposed method. 

\subsection{Pre-Processing Stage}
\label{pre-proc-p2e}

Pre-processing of both the reference single-lead ECG signal and the PPG signal (extracted from the video-PPG data) plays an important role in the efficient training of the P2E-Net. That is, it not only eases the training process but also improves the quality of the reconstructed single-lead ECG waveform. Thus, we first pass the reference ECG signal and the PPG signal through a pre-processing block (see Fig. \ref{fig:preprop}) that is capable of removing baseline, high-frequency noise, and artifacts induced by ambient light from both ECG and PPG signals, and additionally, it de-trends the PPG and ECG signals as well. As a next step, the ECG and PPG signals go to the peaks and valleys detection block where R peaks of ECG are detected during the training phase (see Fig. \ref{fig:ecgpeaksnvalleys}), using the TERMA algorithm \cite{aziz2021ecg}. Afterward, both PPG and reference ECG signals are fed to the synchronization block where the diastolic peak of PPG is synchronized with the R peak of reference ECG.

\begin{figure}
\centering
\includegraphics[width=3.4in]{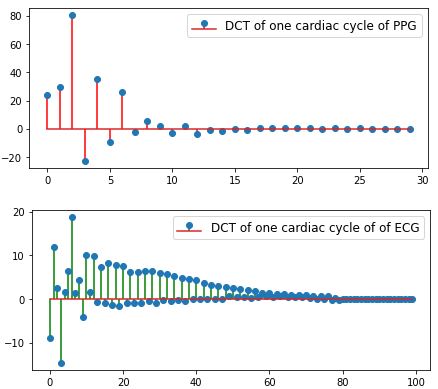}
\caption{DCT coefficients of the PPG signal and the single-lead ECG signal (one cardiac cycle only). The proposed P2E-Net framework learns the regression between the two sets of DCT coefficients.}
\centering
\label{fig:two_DCTs}
\end{figure}

\subsection{The proposed P2E-Net framework}
The proposed approach aims to map one cardiac cycle of vPPG to one cardiac cycle of ECG, as illustrated in Fig. \ref{fig:two_DCTs}. To achieve this, two configurations of P2E-Net are trained and tested, namely, fully connected neural networks and ridge regression models-based configurations. These configurations differ based on the type of input/output layers or the network architecture.
In the Ridge regression configuration, P2E-Net consists of an input layer, a discrete cosine transform (DCT) layer, a regression layer, an inverse DCT (IDCT) layer, and an output layer. On the other hand, in the feedforward neural network (FFNN) configuration, P2E-Net includes the same input, DCT, IDCT, and output layer, but two hidden layers replace the linear regression layer of the Ridge regression configuration. The hidden layers consist of a fully connected layer with 'selu' activation function followed by a batch normalization layer. Both hidden layers and FFNN configuration also apply $L_1$ regularization to avoid overfitting.

\begin{figure}

    \includegraphics[width=0.45\textwidth]{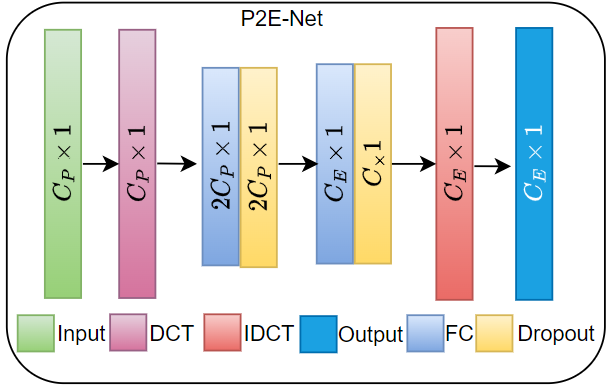}
    \caption{{P2E-Net model architecture in which each fully-connected (FC) layer uses 'tanh' activation and $C_P = 150$ and $C_E = 150$. }}
\label{fig:P2E_Net_mod}
\end{figure}

The input layer of P2E-Net is, actually, the output of pre-processing layer, as described earlier. In the training phase, the pre-processing layer output cycle-wise time-domain vPPG as well as reference/ground truth ECG denoted by $C_P \in \mathcal{R}^{L \times 1}$ and $C_E\in \mathcal{R}^{L \times 1}$ respectively. Then, the cycle-wise time-domain vPPG is fed to the DCT layer of P2E-Net. The DCT layer, then, computes the DCT coefficient $c_P \in \mathcal{R}^{L_P \times 1}$ that corresponds to the cardiac cycle of vPPG. At this stage, the DCT coefficients of the reference ECG cycle $c_E \in \mathcal{R}^{L_E \times 1}$ are also computed offline. The DCT coefficients $c_P$ are fed to the regression/hidden layers, based on the configuration of P2E-Net, that maps them to the DCT coefficients of reconstructed ECG denoted by $\hat{c_E} \in \mathcal{R}^{L_E \times 1}$. The DCT coefficients $c_E$ and $\hat{c_E}$ are used in the loss computation and optimization of the model. Fig. \ref{fig:P2E_Net_mod} provides the complete architecture of the P2E-Net model. 

During the training phase, the iDCT layer remains inactive but once the network is trained and ready to be tested, it activates and  serves to construct the time domain ECG signal from predicted $\hat{c_E}$.The regression model similar to \cite{zhu2019ecg}, when trained, learns a linear mapping from the $c_P$ to the $c_E$. In contrast to regression, P2E-Net in FFNN configuration learns a non-linear mapping from vPPG DCT to the DCT of ECG. In the IDCT layer, the ECG signal is reconstructed.

{\it Training of P2E-Net--}
The P2E-Net is trained for a maximum of 1000 epochs.
All the trainable network parameters are initialized with Xavier initializer. The MAE loss function is used to optimize P2E-Net and for the optimization, Adam optimizer is used.  In the FFNN configuration, a learning rate scheduler, named staircase exponential learning rate decay, with a decay rate of $\exp(0.1)$ is applied to accelerate the convergence and achieve better performance. Moreover, with the aim of avoiding the over-fitting of the model, we apply $L_1$ regularization of hidden layers. An early stopping with verbose 5 is also applied. The best model with the least validation loss is saved after every 5 epochs during training, to see the performance of the best-trained model on test data. Other common training hyperparameters used in these models are shown in Table \ref{tab:table2}. 

\begin{table}
\caption{\label{tab:table2}Values of other hyper-parameters for P2E-Vital-Net. Here $N = 28938$ for pulseDB, $N=8890$ for K20-vPPG, $N=22550$ for BIDMC, and $d = 6380$.}
\begin{center}
\begin{tabular}{|c||*{3}{c|}}
\hline

Hyper-parameters & P2E-Net& Vitals-Net \\\hline\hline
Dataset size & $d$& $N$\\\hline
Validation data set size & $d$& $0.15*N$\\\hline
Test data set size & 0.1*d& $0.10*N$ \\\hline
LESSO data set  & 0.1*d& 5 subjects \\\hline
Training batch size & $100$ &$128$\\\hline
Maximum number of epochs & $1000$ &$1000$\\\hline
Initial learning rate & $10^{-3}$& $10^{-3}$\\\hline
learning rate scheduler & Exponential Decay & -- \\\hline
Scheduler type & Staircase & -- \\\hline
Learning rate decaying factor & exp(0.1)& 0.1\\\hline
Optimizer & Adam & Adam\\\hline

\end{tabular}
\end{center}
 \end{table}

\begin{figure}

    \includegraphics[width=0.45\textwidth]{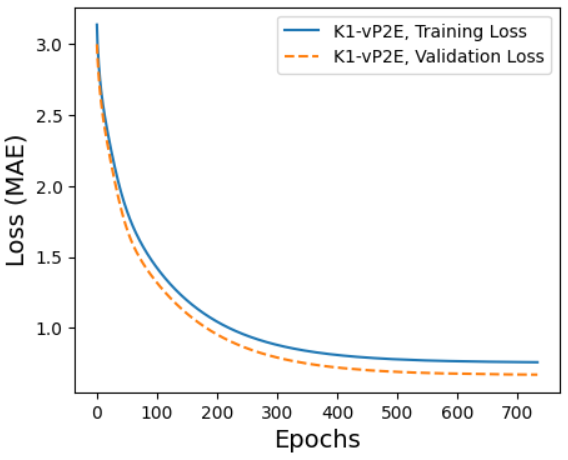}
    \caption{Training and Validation loss of P2E-Net.}
\label{fig:P2E_loss}
\end{figure}

\begin{figure}
    \centering
    \includegraphics[width=0.45\textwidth]
    {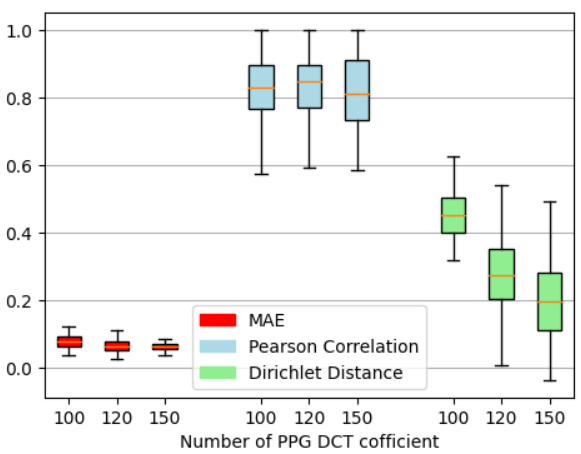}
    \caption{Impact of number of DCT Coefficients on performance of proposed P2E-Net.}
    \label{fig:impct_dct_coeff}
\end{figure}

\begin{figure}
\centering
\begin{subfigure}{0.5\textwidth}
    \includegraphics[width=1\textwidth]{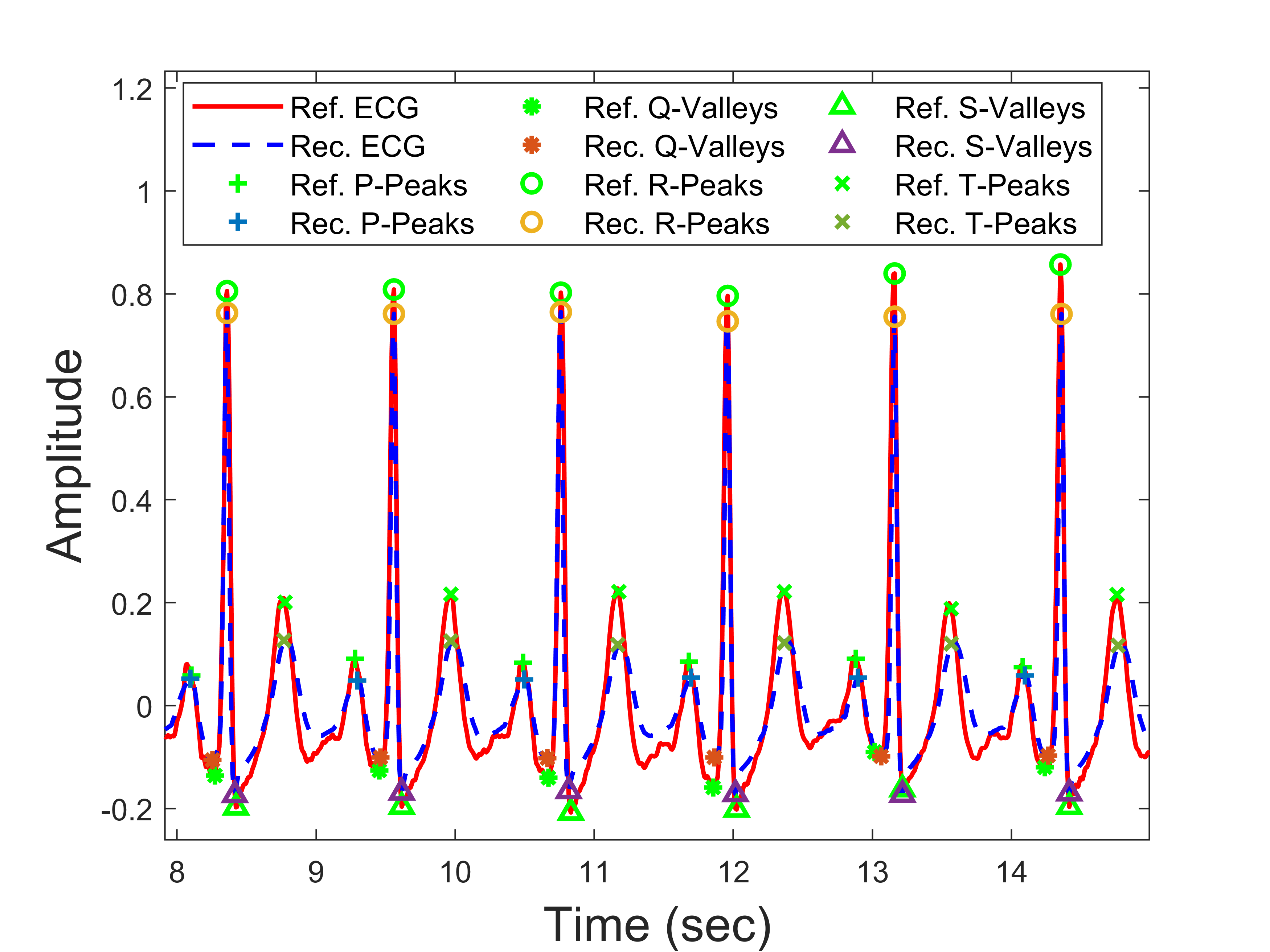}
\end{subfigure}
\vspace{0.0\textwidth}
\caption{"Peaks and Valleys of Reference and Reconstructed ECGs.}
\label{fig:ecgpeaksnvalleys}
\end{figure}

\subsection{Performance Evaluation }

In order to evaluate the performance of P2E-Net, we performed experiments using two different datasets, i.e., we first used the BIDMC dataset and then the vIDEO-P2E dataset. In order to investigate the generalization capability of P2E-Net, we randomly selected 5 subjects that were excluded from the training and validation dataset. From the remaining subjects, we split each session into $80\%$ and $20\%$ for training and validation respectively. Then we  evaluated the performance of the proposed method with left-over data. 
For the rigorous evaluation of P2E-Net, we used three performance metrics, namely Pearson correlation and Dirichlet distance, and MAE. The Pearson correlation coefficient and Dirichlet distance are defined as:
\begin{equation}
P_{corr} = \frac{ (\textbf{x} - \mu_x)^T(\textbf{y} - \mu_y)}{||\textbf{x} - \mu_x||_2 ||\textbf{y} - \mu_y||_2},
\label{eq: loss_function}
\end{equation}
\begin{equation}
l_{dir} = \min(\max_{i \in \mathcal{Q}}(d(x_i, y_i))) , \mathcal{Q} = [1,N]
\label{eq: loss_function}
\end{equation}
In Eq. (2), $\textbf{x}$ represents the reference ECG signal and $\mu_x$ represents its mean. Similarly, $\textbf{y}$ and $\mu_y$ represent the reconstructed ECG signal and its mean, respectively. In Eq. (3), the notation d(*) represents the Euclidean distance between two points. $x_i$ and $y_i$ are the $i^{th}$ elements of \textbf{x} and \textbf{y} respectively and $\mathcal{Q}$ is set of integers $[1, N]$ where $N$ is the length of \textbf{x}.

The P2E-Net model was trained for a maximum of 1000 epochs, but due to early stopping, the training stopped after just over 400 epochs when the MAE loss plateaued at 0.4 for training and 0.46 for validation, as shown in Figure \ref{fig:P2E_loss}. 

We then examined the impact of the number of vPPG DCT coefficients, and Figure \ref{fig:impct_dct_coeff} shows that using a larger number of DCT coefficients improves ECG reconstruction performance in terms of all performance matrices, but not significantly. Overall, the proposed P2E-Net models can generate ECG signals with a mean absolute error below 0.1, a correlation with the reference ECG above 0.8, and a Dirichlet distance around 0.2 on average. It is worth noting that all these results are obtained using left-over datasets, which proves the efficacy of P2E-Net in terms of generalization capability.

\begin{figure*}
     \centering
     \hfill
     \begin{subfigure}[b]{0.32\textwidth}
         \centering
         \includegraphics[width=1.0\textwidth]{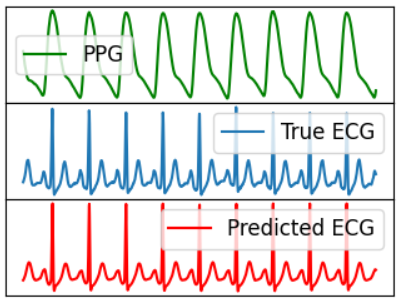}
         \caption{
         }
     \end{subfigure}
     \hfill
     \begin{subfigure}[b]{0.32\textwidth}
         \centering
         \includegraphics[width=1.0\textwidth]{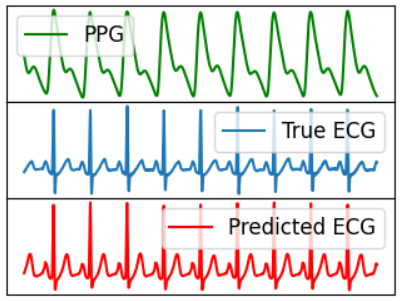}
         \caption{ }
     \end{subfigure}
      \hfill
     \begin{subfigure}[b]{0.32\textwidth}
         \centering
         \includegraphics[width=1.0\textwidth]{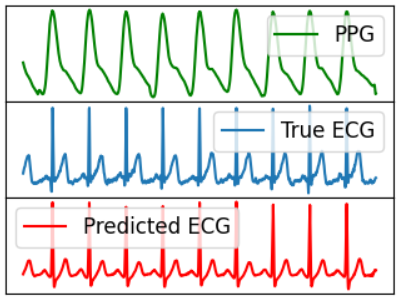}
         \caption{}
     \end{subfigure}
     \hfill

          \centering
     \hfill
     \begin{subfigure}[b]{0.32\textwidth}
         \centering
         \includegraphics[width=1.0\textwidth]{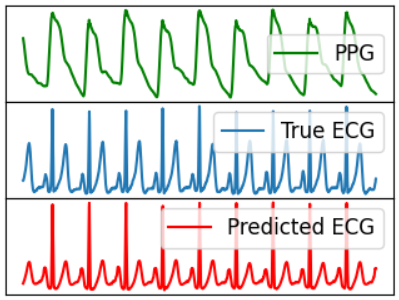}
         \caption{ 
         }
     \end{subfigure}
     \hfill
     \begin{subfigure}[b]{0.32\textwidth}
         \centering
         \includegraphics[width=1.0\textwidth]{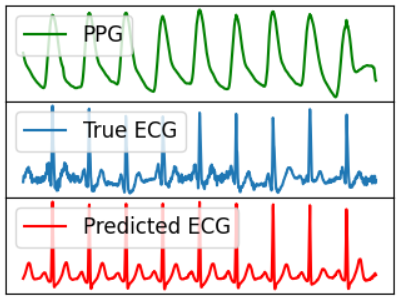}
         \caption{ 
         }
     \end{subfigure}
      \hfill
     \begin{subfigure}[b]{0.32\textwidth}
         \centering
         \includegraphics[width=1.0\textwidth]{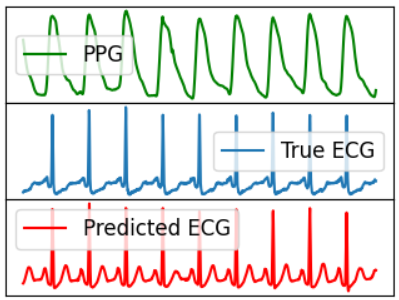}
         \caption{}
     \end{subfigure}
     \hfill
    \caption{Some ECG reconstruction samples using the {ridge regression} method. }
    \label{fig:P2E-rig}
\end{figure*}

\begin{figure*}
     \centering
     \hfill
     \begin{subfigure}[b]{0.32\textwidth}
         \centering
         \includegraphics[width=1.0\textwidth]{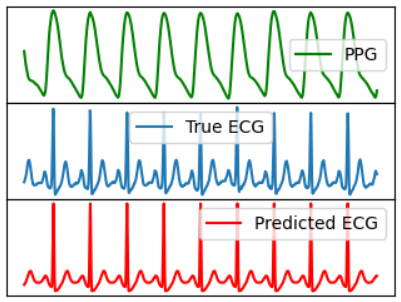}
         \caption{
         }
     \end{subfigure}
     \hfill
     \begin{subfigure}[b]{0.32\textwidth}
         \centering
         \includegraphics[width=1.0\textwidth]{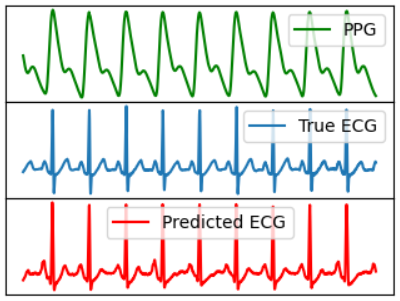}
         \caption{ }
     \end{subfigure}
      \hfill
     \begin{subfigure}[b]{0.32\textwidth}
         \centering
         \includegraphics[width=1.0\textwidth]{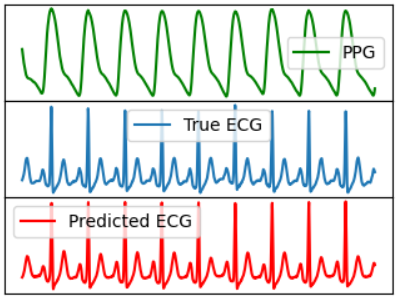}
         \caption{}
     \end{subfigure}
     \hfill

          \centering
     \hfill
     \begin{subfigure}[b]{0.32\textwidth}
         \centering
         \includegraphics[width=1.0\textwidth]{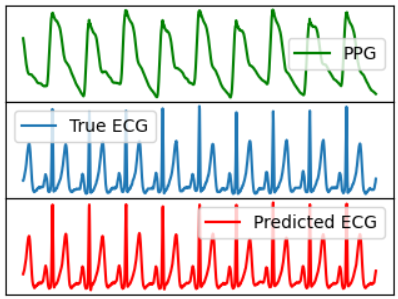}
         \caption{ 
         }
     \end{subfigure}
     \hfill
     \begin{subfigure}[b]{0.32\textwidth}
         \centering
         \includegraphics[width=1.0\textwidth]{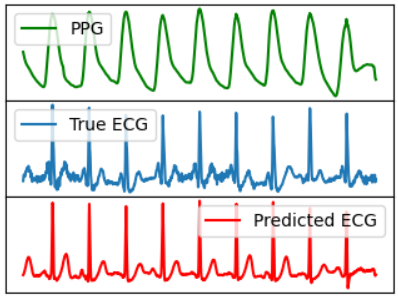}
         \caption{ 
         }
     \end{subfigure}
      \hfill
     \begin{subfigure}[b]{0.32\textwidth}
         \centering
         \includegraphics[width=1.0\textwidth]{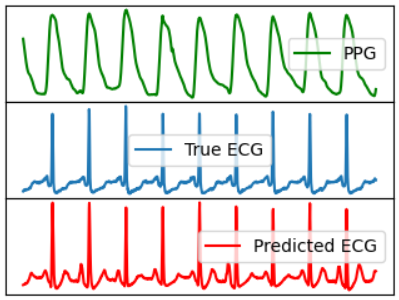}
         \caption{}
     \end{subfigure}
     \hfill
    \caption{Some ECG reconstruction samples using proposed {P2E-Net} model. }
    \label{fig:P2E-Net}
\end{figure*}

As ECG signals are mainly characterized by the P, QRS, and T peaks, we performed a cardiac cycle-level investigation. We selected the ECG signals of five subjects from the left-over data and detected all peaks as shown qualitatively in Fig \ref{fig:ecgpeaksnvalleys}. The peak-level quantitative performance of P2E-Net is presented in Table \ref{tab:4}, which shows that neural network-based models outperform the ridge regression model, and both methods efficiently reconstruct the ECG. Finally, Figs. \ref{fig:P2E-rig},\ref{fig:P2E-Net} show the qualitative performance of the ridge regression method and FFNN-based method whereby the reconstructed ECG waveforms show a high morphological similarity with the reference ECG waveform (for a few chosen subjects). 

In summary, the P2E-Net efficiently reconstructs ECG signals with a shallow FFNN model, making it highly suitable for deployment on smartphones. 


\begin{table*}
\label{tab:table5}
\caption{ Performance of P2E-Net. Mean and std. dev. of Peaks and Valleys of reconstructed ECG waveforms for 5 subjects.
}
\label{tab:4}
\begin{center}
\begin{tabular}{|c||*{10}{c|}}
\hline

Method &\multicolumn{2}{c|}{P-Peaks}  & \multicolumn{2}{c|}{Q-Valleys}  &\multicolumn{2}{c|}{R-Peaks}  & \multicolumn{2}{c|}{S-Valleys} & \multicolumn{2}{c|}{T-Peaks}   \\\hline

~ & MMAE & MSAE &MMAE & MSAE  &MMAE& MSAE & MMAE & MSAE &MMAE & MSAE   \\\hline

DCT+Ridge regression    & $0.0487$ & $0.01996$   & \textbf{0.05722} & 0.0243 &  0.0883 & 0.0419 &  \textbf{0.0850}
 &   \textbf{0.0291}    &0.1125 & 0.0319\\\hline
 
DCT+FFNN      & \textbf{0.0418}&	\textbf{0.0177}&	0.0612 &	\textbf{0.0177}&	\textbf{0.059}&	\textbf{0.0381}&	0.09516&	\textbf{0.0303}&	\textbf{0.1044}&	\textbf{0.0275}\\\hline


\end{tabular}
\end{center}
\end{table*}

\section{Conclusion}
\label{conc}
This work demonstrated the feasibility of using a smartphone as an initial diagnostic tool to measure one's body vitals, i.e., pulse rate, SpO2 and respiratory rate, and a single-lead ECG. A number of custom-built CNNs and FFNNs (including a vision transformer and a CLIP model) were implemented to extract the body vitals as well as the single-lead ECG from the video-PPG signal recorded from rear camera of the smartphone. This work invites smartphone manufacturers and android app developers to deliberate and standardize algorithms to measure body vitals and single-lead ECG so that people living in remote and far-away areas may benefit from it by keeping track of their well-being using their smartphones.

\footnotesize{
\bibliographystyle{IEEEtran}
\bibliography{main}
}
\vfill\break
\end{document}